\documentclass{scspaperproc}
\usepackage{textcase}
\usepackage{latexsym}
\usepackage{graphicx}
\usepackage{mathptmx}

\usepackage{amsmath}
\usepackage{amsfonts}

\usepackage{amssymb}
\usepackage{amsbsy}
\usepackage{amsthm}
\usepackage{algorithm}
\usepackage{algpseudocode}
\usepackage{tikz}
\usetikzlibrary{arrows.meta}
\usepackage{placeins}
\usepackage{float}

\usepackage{hyphenat}
\newtheoremstyle{scsthe}
{8pt}
{8pt}
{\it}
{}
{\bf}
{.}
{.5em}
{}

\theoremstyle{scsthe}

\usepackage[english]{babel}

\IfFileExists{accessibility.sty}{\usepackage{accessibility}}{}

\makeatletter
\@ifpackageloaded{hyperref}{%
  \AtBeginDocument{\hypersetup{
    colorlinks=true,
    linkcolor=black,
    citecolor=black,
    urlcolor=blue,
    pdfborder={0 0 0}
  }}%
}{%
  \usepackage[
    colorlinks=true,
    linkcolor=black,
    citecolor=black,
    urlcolor=blue,
    pdfborder={0 0 0}
  ]{hyperref}%
}
\makeatother

\hypersetup{
  pdftitle={ANNSIM 2026 Proceedings Paper},
  pdfauthor={Proceedings Authors},  
  pdfsubject={ANNSIM 2026 Conference},
  pdfkeywords={}
}

\begin{document}



\SCSpagesetup{Bishi, Bowman, and Miller}

\def\SCSconferencename{Annual Modeling and Simulation Conference}

\def\SCSconferenceacro{ANNSIM'26}

\def\SCSpublicationyear{2026}

\def\SCSconferenceeditors{G. Rabadi, V. Prabhu, R. Cárdenas, A. Bany Abdelnabi, J. Jabbour, and M. Germanos}

\def\SCSconferencedates{May 4-7}

\def\SCSconferencevenue{University of Central Florida, Orlando, Florida, USA}


\title{BEYOND CORRIDOR AVERAGES: LANE-LEVEL VALIDATION OF MICROSCOPIC FREEWAY SIMULATION WITH EMPIRICALLY PARAMETERIZED ARRIVALS
}


\author[\authorrefmark{1}]{Korede R. Bishi}
\author[\authorrefmark{2}]{Casey Bowman}
\author[\authorrefmark{1}]{John A. Miller}

\affil[\authorrefmark{1}]{School of Computing, University of Georgia, Boyd Research and Education Center, Athens, GA 30602, USA}
\affil[\authorrefmark{2}]{Department of Mathematics, University of North Georgia, 82 College Circle, Dahlonega, GA 30597, USA}

\maketitle
\section*{Abstract}
Traffic flow modeling is an essential part of civil planning around the world, and traffic simulation is an
important component of the analyses that city planners must perform to ensure safe and efficient road networks for
their citizens. This work advances microscopic traffic simulation through lane‑level validation of speed and flow
dynamics, addressing a critical gap in conventional macro‑aggregated approaches. We systematically evaluate the
trade‑offs between numerical integration methods and vehicle arrival processes when reproducing empirical trajectory
data from California’s freeway network. Results demonstrate that numerical‑integrator choice has negligible impact
on simulation accuracy, with less than one percent variation across eight methods, while vehicle arrival‑process modeling substantially affects fidelity. The shifted Erlang‑2 distribution reduces flow‑prediction error by approximately
28\% compared to Poisson processes by enforcing a realistic minimum headway. Lane‑level validation across five
detector stations reveals lane‑specific dynamics that aggregated metrics often obscure.

\textbf{Keywords: }{Lane-level validation, Vehicle headway distribution, Numerical integration, Loop detector data, Arrival process}


\section{Introduction}
\label{sec:intro}
Every day, citizens in large cities around the world struggle with inefficient and unsafe traffic conditions.
There remains a need for traffic analysis that can improve the safety and efficiency of highways and suburban roads that carry millions of commuters to work each day. Such analysis can benefit greatly from accurate simulation models \cite{law2007simulation}.While algorithms such as deep learning networks can help solve problems related to current data, accurate simulation models are better suited to answer what‑if questions, such as physical road changes or major traffic distruptions from natural disasters \cite{fishman2001discrete}.

The accuracy of traffic simulation models depends on several components of the model.
One primary concern is developing an accurate vehicle‑arrival model. In this work, we highlight how the use of a shifted Erlang distribution improves on other arrival models \cite{cowan1975useful, li2017vehicle}.
Car‑following models are systems of differential equations that require numerical solvers to compute updates to vehicle speeds and positions. In this work, we offer several solutions that span a spectrum of trade‑offs - between model accuracy and computational efficiency \cite{treiber2015comparing, pvrikryl2017comparing}.

Our investigation yields three key findings:
First, we demonstrate that the choice of numerical integrator has negligible impact on the simulation accuracy for microscopic discrete‑event traffic simulation using car‑following models. We evaluated eight numerical solvers, ranging from simple ballistic kinematics to higher‑order Runge‑Kutta methods. Across all integrators, the variation in the goodness‑of‑fit metrics remained below one-percent, indicating that increasing numerical sophistication provides little benefit in prediction accuracy. These findings are consistent with prior comparative studies \cite{treiber2015comparing, pvrikryl2017comparing}.

Second, we find that the choice of vehicle‑arrival process has a substantially greater effect on simulation fidelity
than the numerical‑integration scheme. Specifically, the shifted Erlang‑2 arrival distribution reduces overall error compared to the best‑performing Poisson‑based configuration. This improvement was consistent across all numerical‑integrator choices, highlighting the relative importance of arrival‑process modeling.

Third, we validate our simulation (both flow and speed) against empirical lane-level data from the Caltrans Performance Measurement System (PeMS) \cite{caltran_pems}, using observations from five mainline loop detectors across a four‑lane freeway segment, supplemented by two on‑ramp detector stations.

\section{BACKGROUND AND RELATED WORK}
\label{sec:related}
\subsection{Car-following Models}
Microscopic traffic simulation relies on car‑following models to describe the behavior of vehicles as they follow
one another in traffic. Numerous models have been proposed, falling into several broad categories, including
stimulus‑response models, optimal‑velocity models, collision‑avoidance models, continuous‑time models,
psycho‑physical models, and cellular‑automaton models. A comprehensive genealogy of car‑following model evolution
was presented by \cite{van2015genealogy}; a comprehensive treatment of traffic flow dynamics appears in \cite{kesting2013traffic}. The most widely used continuous‑time CF model today is the Intelligent
Driver Model (IDM) \cite{treiber2000congested}. It has performed well in comparative studies such as
\cite{zhang2021comprehensive, pourabdollah2017calibration}, which both found that IDM outperforms other models.

\subsection{Numerical Integration in Traffic Simulation}
Car‑following models typically provide a formula to calculate either the new acceleration or velocity of each
vehicle in the traffic system. This calculation is usually based on the relative distances, speeds, and
accelerations of the vehicle in focus and its immediate leader. When the model calculates a new acceleration for a
vehicle, a numerical integrator must solve the differential equation to compute first the new velocity and then the
distance traveled. Many numerical integrators have been used, with the key consideration being a balance between
accuracy and computational efficiency. Treiber and Kanagaraj compare Euler’s Method, Heun’s (Trapezoidal) Method, a
fourth‑order Runge‑Kutta (RK4) method, and an update method based on Newtonian ballistics
\cite{treiber2015comparing}. They find that the ballistic update method performs well in all cases, particularly
when dealing with stop‑and‑start traffic and lane changing. P\v{r}ikryl and Vani\v{s} reported similar results,
though found Euler’s method to perform somewhat more strongly \cite{pvrikryl2017comparing}. Both studies found that
the ballistic update based on simple Newtonian equations performs more accurately and more efficiently than other
methods.
\subsection{Arrival Processes}
Proper modeling of vehicle headways is essential in traffic simulation. Vehicles must enter the simulation horizon
in a realistic manner. Poisson distributions are still used extensively; however, they have been shown to be inappropriate in complex traffic scenarios with multiple busy periods of varying intensities \cite{meng2009self}.
Other models have been proposed for these scenarios, including those based on the gamma distribution or closely
related distributions such as the Erlang and Pearson families \cite{li2017vehicle}. Various comparative studies of
different headway models include Li and Chen \cite{li2017vehicle}, Roy and Saha \cite{roy2018headway}, and Singh et
al. \cite{singh2020multivariate}.

\subsection{Lane-Level Validation}
Vehicle‑arrival processes and car‑following models have been widely studied in the
traffic‑simulation literature; however, validation and reporting are most often performed using aggregated flow
measures, as illustrated in \cite{11118854}. While such approaches provide useful corridor‑level insights, they can obscure lane‑specific speed and flow dynamics. Our work differs from \cite{11118854} in three key aspects: consideration of numerical‑integrator choice, explicit arrival‑process modeling, and systematic lane‑level validation of both speed and flow. Specifically, \cite{11118854} focuses on a single bottleneck loop detector and evaluates lane‑wise speed and flow using two car‑following models.

In contrast, we perform lane‑level validation across five detector stations with two on‑ramp merge locations, reporting speed and flow dynamics for all lanes at regular time intervals. To our knowledge, no prior work has reported a comprehensive lane-level analysis of microscopic simulation accuracy across multiple detector stations and lanes at this spatial and temporal resolution.

\section{Methodology}
\label{sec:methodology}
Our aim in this work is to design a simulation setup that reports not only macro-level validation metrics but also micro-level vehicle dynamics at fixed temporal intervals. Achieving reliable micro-level accuracy requires
a car-following model whose acceleration formulation responds smoothly to changes in speed, spacing, and relative velocity.

For this reason, we adopt IDM model \cite{treiber2000congested}. IDM defines acceleration as a continuous function of the current velocity, net distance gap, and approach rate, allowing it to capture gradual transitions between free-flow, following, and braking regimes. As emphasized in the original formulation \cite{treiber2000congested}, Unlike the Gipps model \cite{gipps1981behavioural}, which introduces abrupt acceleration and deceleration behavior in stop-and-go regime. For arrival processes, we experimented with two approaches: the Poisson process and Shifted Erlang‑2 arrivals.
Although Krauss which is the default in Simulation of Urban Mobility (SUMO) \cite{lopez2018microscopic}, IDM tends to be better for physical realism and calibration.

\subsection{Intelligent Driver Model}

The IDM model computes the longitudinal acceleration
$a_n$ of vehicle $n$ following a predecessor $p$ as shown in Equation~\eqref{eq:idm-accel}  \cite{treiber2000congested}:

\begin{equation}
\label{eq:idm-accel}
a_n = a_{\max} \left[
1 - \left( \frac{v_n}{v_{\max}} \right)^\delta
- \left( \frac{s^*(v_n, \Delta v)}{s_n} \right)^2
\right],
\end{equation}
where $v_n$ is the velocity of vehicle $n$, $\Delta v = v_n - v_p$ is the
relative velocity to the leader, and $s_n$ is the bumper-to-bumper gap.
The model's target spacing $s^*$ is given by:
\begin{equation}
\label{eq:idm-spacing}
s^*(v_n, \Delta v) =
s_0 + \max\!\left(0,\, v_n T + \frac{v_n \Delta v}{2\sqrt{a_{\max} b}} \right).
\end{equation}

The IDM's behavior is sensitive to five parameters: minimum gap $s_0$, maximum acceleration $a_{\max}$, comfortable deceleration $b$, desired time headway $T$, and desired velocity $v_{\max}$.
Following Equation~\eqref{eq:idm-accel} and \eqref{eq:idm-spacing}, we set $s_0 = 5.0$ m, $a_{\max} = 4.0$ m/s$^2$, $b = 2.0$ m/s$^2$, $T = 3.0$ s; these are applied uniformly across all lanes, while $v_{\max}$ is set per lane and interval from observed PeMS speed data.

\subsection{Coupled ODE Formulation}
IDM acceleration depends on both position and velocity simultaneously: position determines the gap $s_n = x_p - x_n$, and velocity determines the approach rate $\Delta v = v_n - v_p$. To capture this coupling, we formulate the vehicle state as a vector (Equation~\eqref{eq:state-vector}):
\begin{equation}
\label{eq:state-vector}
\mathbf{y}_n = \begin{bmatrix} x_n \\ v_n \end{bmatrix}
\end{equation}
At each simulation step, the state update is governed by a couple initial value problem (Equation~\eqref{eq:coupled-ode}):
\begin{equation}
\label{eq:coupled-ode}
\frac{d\mathbf{y}_n}{dt} = \begin{bmatrix} v_n \\ a_{\text{IDM}}(x_n, v_n; \tilde{x}_p, \tilde{v}_p) \end{bmatrix}
\end{equation}
where $(\tilde{x}_p, \tilde{v}_p)$ are the leader's state values, frozen at the start of each integration step, which we simulated using \textbf{Algorithm 1} in the
\textbf{Appendix}. $a_{\text{IDM}}$ is the IDM acceleration from Equation~\eqref{eq:idm-accel}.

\subsection{Shifted Erlang-2 Arrivals}

Once vehicle dynamics were integrated using \textbf{Algorithm~1}, the remaining challenge was generating vehicle arrivals at realistic inter-arrival rates. Vehicle arrivals are commonly modeled using exponential inter-arrival times, corresponding to a Non-Homogeneous Poisson Process (NHPP) \cite{law2007simulation}. However, exponential distributions exhibit high variance producing irregular headway that poorly represent steady traffic flow.

To reduce headway variability, we adopt a Non-Homogeneous Erlang process, where Erlang-2 headways are modeled as the sum of two exponential stages, yielding smoother arrivals than exponential headways. A shifted Erlang model further adds a minimum headway $\tau$ enforcing a hard lower bound on spacing consistent with finite vehicle length and driver reaction constraints \cite{cowan1975useful}. Following this approach, we employ a shifted Erlang-2 distribution, which preserves the smoothness of the Erlang family while enforcing a minimum feasible headway.

\textbf{Variate generation.} Let $U_1, U_2 \sim \text{Uniform}(0,1)$ be independent uniform variates, $\mu > 0$ denote the mean of each exponential stage, and $\tau > 0$ the minimum headway. An Erlang-2 variable is the sum of two independent exponentials: $X = E_1 + E_2$, where each $E_i = -\mu \ln(U_i)$ via the logarithmic transformation. This gives $X = -\mu \ln(U_1 U_2)$. Adding the shift $\tau$ yields the headway:
\begin{equation}
\label{eq:erlang-variate}
Y = \tau - \mu \ln(U_1 U_2), \quad \text{with mean } \mathbb{E}[Y] = \tau + 2\mu.
\end{equation}

\textbf{Data-driven parameterization.}
A mean inter-arrival time is computed for each lane $\ell$ and for each interval 15 minute time interval $t$ is $\bar{\mu}_{\ell,t} = \Delta T / N_{\ell,t}$, where $\Delta T = 900$ seconds and $N_{\ell,t}$ is the observed count. Setting $\mathbb{E}[Y] = \bar{\mu}_{\ell,t}$ and solving:
\begin{equation}
\label{eq:erlang-param}
\mu = \frac{\bar{\mu}_{\ell,t} - \tau}{2}.
\end{equation}

\section{SIMULATION SETUP}
\label{sec:setup}
\subsection{Study Corridor}
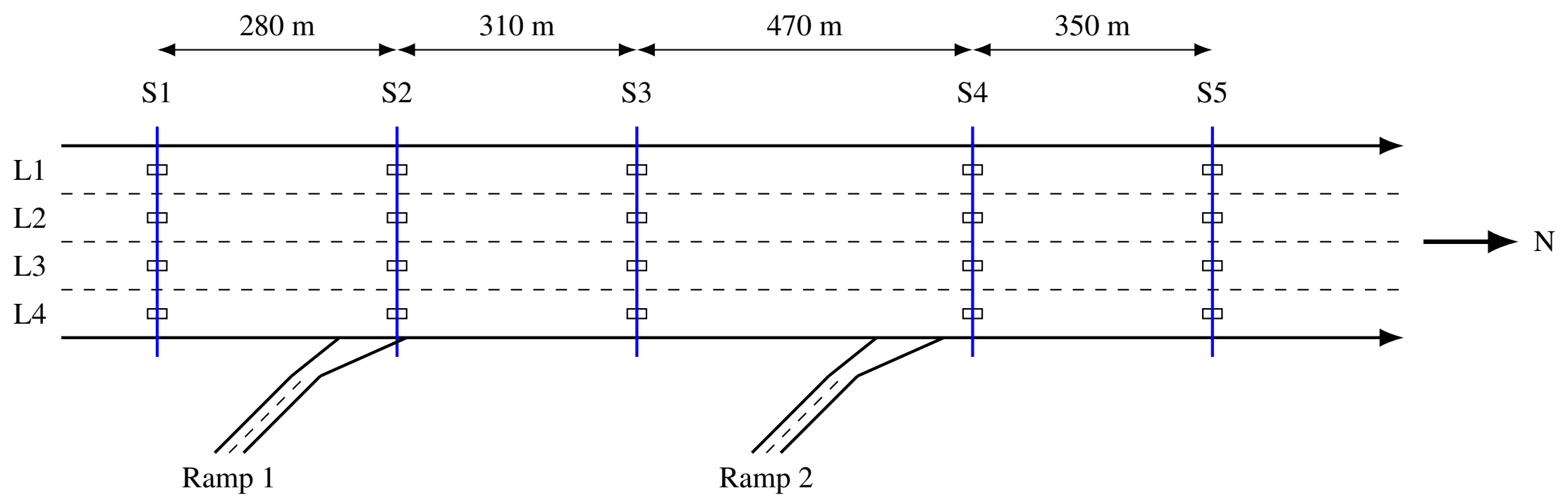
\begin{figure}[H]
\centering
\begin{tikzpicture}[scale=0.9, >=Latex]
\draw[thick,->] (0,0) -- (14,0);     
\draw[thick,->] (0,2.0) -- (14,2.0); 

\foreach \y in {0.5,1.0,1.5} {
    \draw[dashed] (0,\y) -- (14,\y);
}

\node[left] at (0,1.75) {\scriptsize L1};
\node[left] at (0,1.25) {\scriptsize L2};
\node[left] at (0,0.75) {\scriptsize L3};
\node[left] at (0,0.25) {\scriptsize L4};


\foreach \x in {1,3.5,6,9.5,12} {
    \foreach \y in {1.75,1.25,0.75,0.25} {
        \draw[fill=white] (\x-0.1,\y-0.05) rectangle (\x+0.1,\y+0.05);
    }
}

\foreach \x/\lab in {1/S1,3.5/S2,6/S3,9.5/S4,12/S5} {
    \draw[blue, thick] (\x,-0.2) -- (\x,2.2);
    \node[above] at (\x,2.3) {\scriptsize \lab};
}

\draw[thick] (1.6,-1.2) -- (2.4,-0.4);
\draw[thick] (1.9,-1.2) -- (2.7,-0.4);
\draw[dashed] (1.75,-1.2) -- (2.55,-0.4);
\draw[thick] (2.4,-0.4) -- (2.9,0.0);
\draw[thick] (2.7,-0.4) -- (3.6,0.0);

\node[below] at (1.75,-1.2) {\scriptsize Ramp 1};

\draw[thick] (7.2,-1.2) -- (8.0,-0.4);
\draw[thick] (7.5,-1.2) -- (8.3,-0.4);
\draw[dashed] (7.35,-1.2) -- (8.15,-0.4);
\draw[thick] (8.0,-0.4) -- (8.5,0.0);
\draw[thick] (8.3,-0.4) -- (9.2,0.0);

\node[below] at (7.35,-1.2) {\scriptsize Ramp 2};

\draw[<->] (1,3.0) -- (3.5,3.0);
\node[above] at (2.25,3.0) {\scriptsize 280 m};

\draw[<->] (3.5,3.0) -- (6,3.0);
\node[above] at (4.75,3.0) {\scriptsize 310 m};

\draw[<->] (6,3.0) -- (9.5,3.0);
\node[above] at (7.75,3.0) {\scriptsize 470 m};

\draw[<->] (9.5,3.0) -- (12,3.0);
\node[above] at (10.75,3.0) {\scriptsize 350 m};

\draw[very thick,->] (14.2,1.0) -- (15.2,1.0);
\node[right] at (15.2,1.0) {\scriptsize N};
\end{tikzpicture}
\vspace{-1.0em}
\caption{Study corridor: US-101 northbound, Donald Doyle Highway segment, San Francisco. Five mainline loop detector stations (S1--S5) monitor four lanes. Two on-ramps merge between S1--S2 and S3--S4.}
\label{fig:corridor}
\end{figure}

We model a 1.4 km segment of northbound US-101 in the Donald Doyle Highway area  near Danville, California (Figure~\ref{fig:corridor}). The corridor features four mainline lanes with two on-ramp merge points. Five loop detector stations positioned along the mainline at approximately 280 m, 310 m, 470 m, and 350 m intervals, provides flow and speed measurements for our experimentation.
The physical layout follows Figure~\ref{fig:corridor}. This corridor exhibits recurring congestion
during morning and evening peak periods, making it suitable for validating car-following model calibration at the micro level.

\subsection{Empirical Data}
Traffic data were obtained from the PeMS \cite{caltran_pems}, which archives data from loop-detector observations from California’s freeway network. We collected data from five mainline detectors and two on-ramp detectors along the study corridor, aggregated into 15-minute intervals (6:00 AM to 6:00 PM, 48 intervals).
Each mainline detector provides lane-level measurements across four lanes, yielding flow and speed time series for validation. On-ramp demand is derived from two single-lane ramp detectors as shown in Figure~\ref{fig:corridor}.
\begin{figure}[H]
    \centering
    \includegraphics[width=\linewidth]{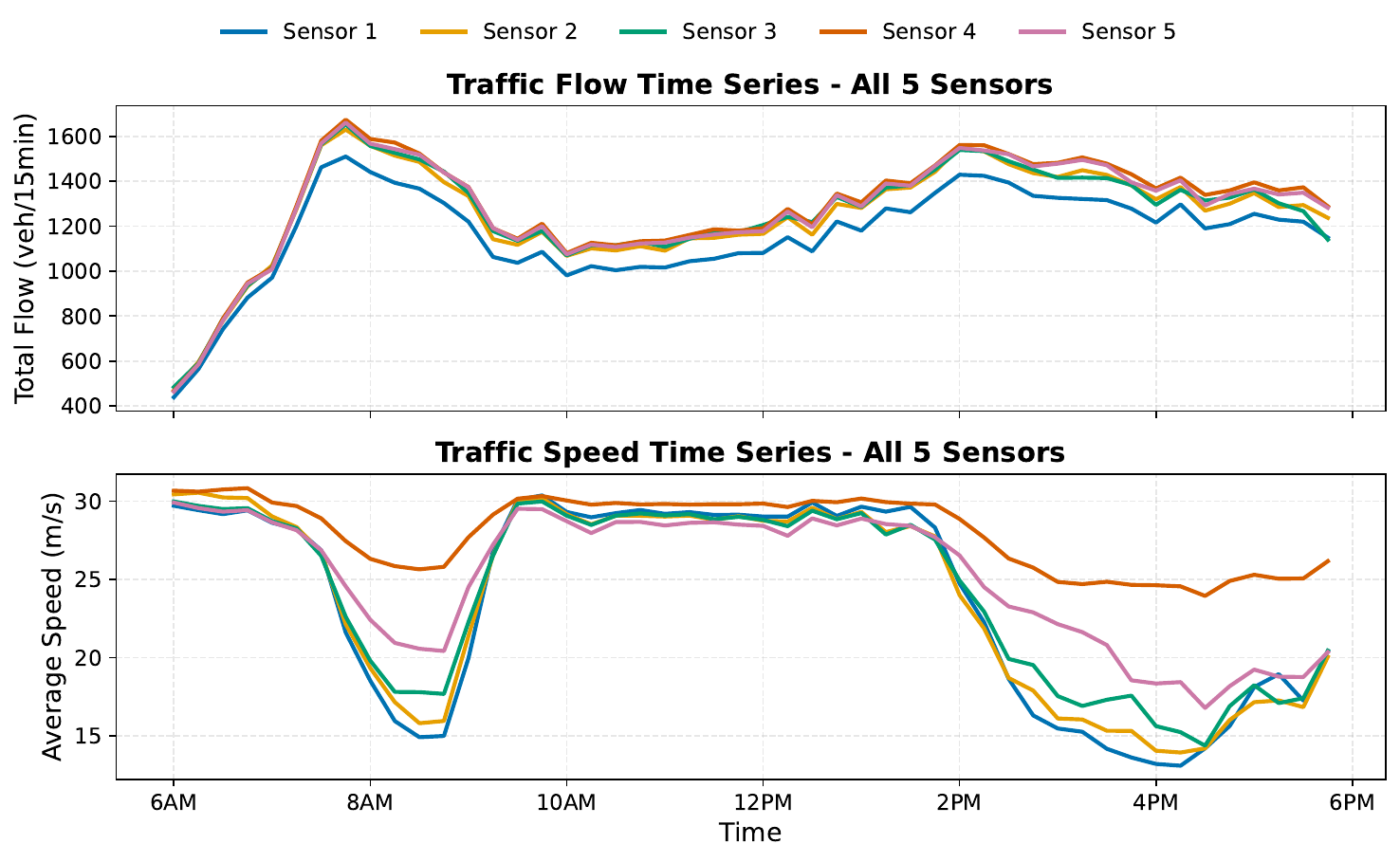}
    \vspace{-2.0em}
    \caption{Observed flow (top) and speed (bottom) at five mainline sensors (S1--S5) along the US-101 corridor. Daily volumes range from 56,102 to 62,445 vehicles; average speeds from 23.6 to 28.0 m/s.}
    \label{fig:sensor_timeseries}
\end{figure}
We selected a stretch of road to ensure sufficient data quality based upon PeMS self reported detector health and data fidelity \cite{law2007simulation, sargent2020verification}. Sensor 4 lanes 2--3 from Figure~\ref{fig:sensor_timeseries} exhibited anomalous speed readings (likely detector malfunction). Such data quality issues are common in loop-detector datasets \cite{li2020trajectory}.

 \subsection{Simulation Model}
\begin{figure}[H]
    \centering
    \includegraphics[width=0.90\linewidth]{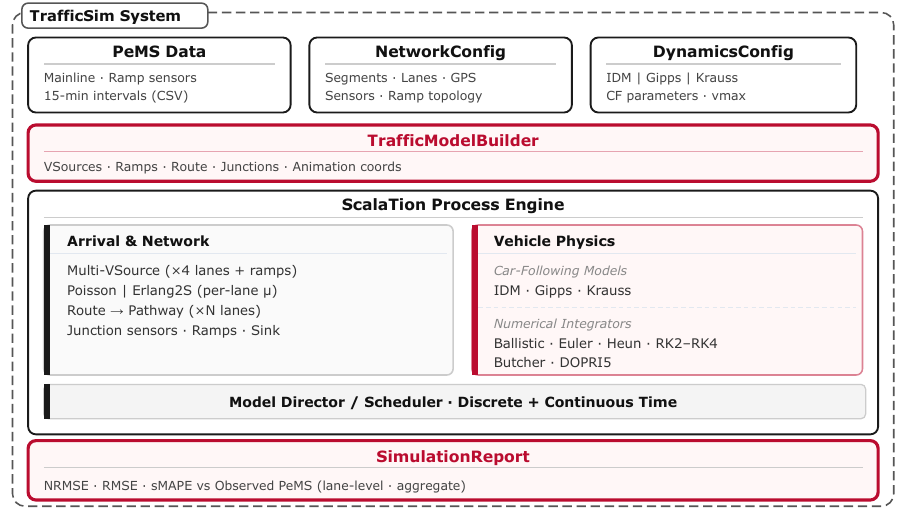}
    \vspace{-0.5em}
    \caption{ PeMS data parameterized per-lane arrival sources assembled by the TrafficModelBuilder into the ScalaTion 2.0 hybrid discrete–continuous simulation engine}
    \label{fig:architecture}
\end{figure}

The simulation is implemented in ScalaTion 2.0 \cite{miller2010using, miller2017advanced}, an open-source Scala-based modeling and simulation framework. The framework supports multiple modeling paradigms including process-based and event-based approaches. Our microscopic traffic simulation combines discrete-event and discrete-time utilizing a process-oriented paradigm in which each vehicle is an independent coroutine-based actor \cite{miller2017advanced}.
The simulation clock runs for 43,200 seconds (12 hours), with vehicles entering according to the time-varying arrival schedule (Figure~\ref{fig:architecture}).


\section{RESULTS AND VALIDATION}
\label{sec:results}
We validate simulation performance using multiple goodness-of-fit metrics, including Sum of Squared Errors (SSE), Mean Absolute Error (MAE), Root Mean Square Error (RMSE), Normalized RMSE (NRMSE), and symmetric Mean Absolute Percentage Error (sMAPE) for both flow and speed at each sensor and lane.

\enlargethispage{2\baselineskip}
\subsection{Numerical Integrator Result}
Our experiments indicate that numerical integrator choice has limited impact on lane-level flow and speed accuracy for the IDM car-following model. While small differences are observable, their magnitude remains within $\sim 1\%$ across repeated stochastic runs, suggesting that integration error is not a primary driver of simulation outcomes in this setting. In contrast, runtime differences are substantial and consistent. From a practical simulation standpoint, this makes computational cost the dominant factor when selecting an integrator. A similar conclusion was reached by \cite{treiber2015comparing, pvrikryl2017comparing}. Under this criterion, the ballistic method is clearly preferable, achieving an average total runtime of \textbf{17 minutes and running approximately $\sim$ 3X} faster than Runge–Kutta fourth order (RK4) and Dormand-Prince (DOPRI5) \cite{dormand1980family}. Runtime results for all integrators are reported in Table~\ref{tab:integrator-comparison}. Speedup relative to Ballistic.
\begin{table}[H]
    \centering
    \begin{tabular}{lccc}
        \hline
        \textbf{Integrator} & \textbf{Duration (min)} & \textbf{Fitness} & \textbf{Speedup} \\
        \hline
        Ballistic & \textbf{17.5} & \textbf{15.67} & 1.00$\times$ \\
        RK4       & 50.9 & 15.72 & 0.34$\times$ \\
        RK3       & 48.0 & 15.74 & 0.36$\times$ \\
        RK2       & 48.4 & 15.75 & 0.36$\times$ \\
        Heun      & 44.8 & 15.78 & 0.39$\times$ \\
        Euler     & 44.3 & 15.79 & 0.40$\times$ \\
        DOPRI5    & 51.1 & 15.82 & 0.34$\times$ \\
        \hline
    \end{tabular}
    \begin{flushleft}
        \footnotesize
        \textit{Note:} Fitness $= 0.5 \times$ Flow\_NRMSE $+ 0.5 \times$ Speed\_NRMSE.
    \end{flushleft}
    \caption{Numerical integrator comparison: Runtime (for the entire simulation) vs.\ fitness (Erlang2S arrivals). NRMSE values shown as percentages. All integrators achieve $<$1\% fitness variation. }
     \label{tab:integrator-comparison}
\end{table}
\subsection{Arrival Results}
\begin{figure}[H]
  \centering
  \includegraphics[width=\textwidth]{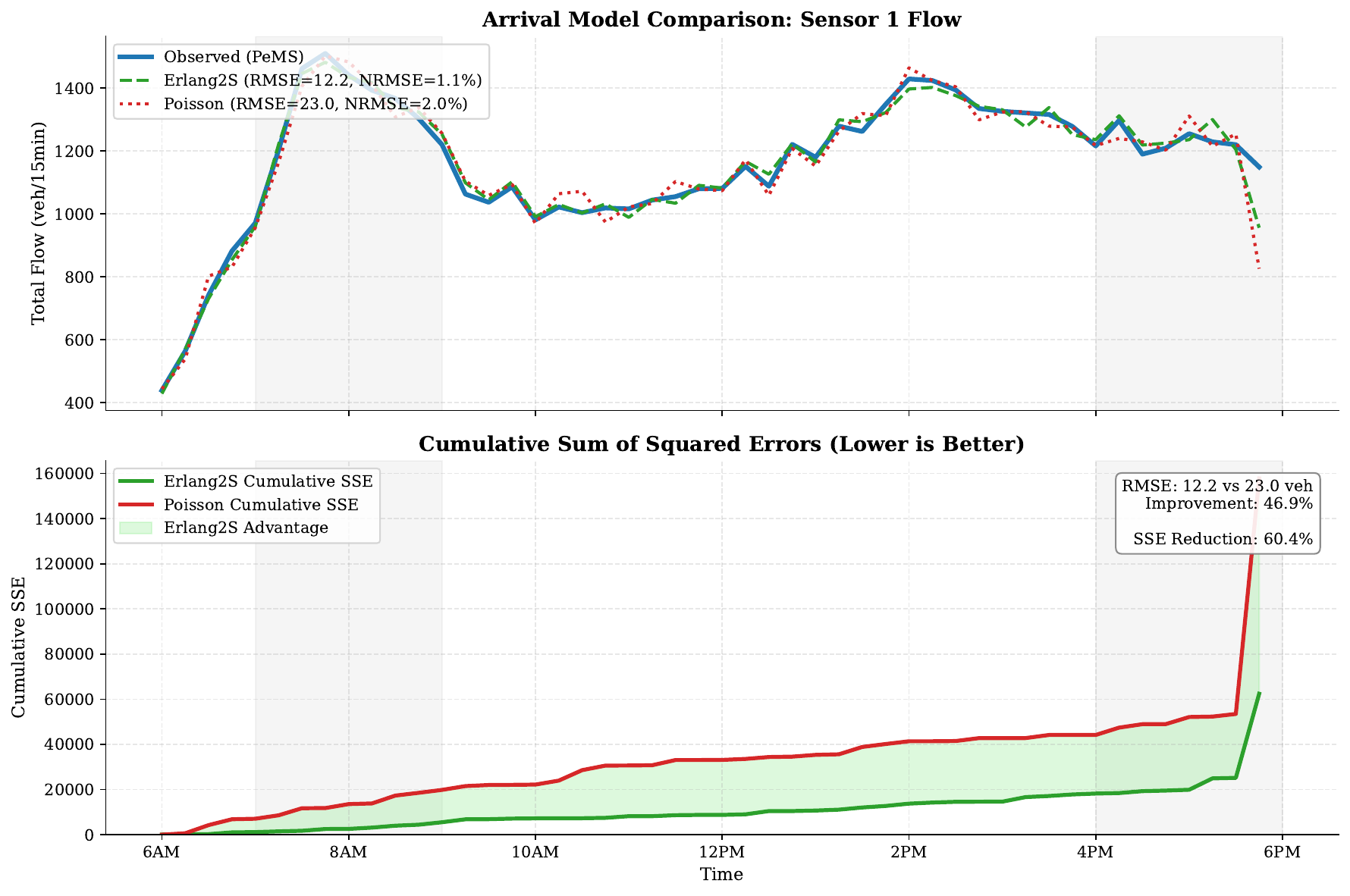}
  \vspace{-2.5em}
  \caption{Cumulative flow prediction error at Sensor~1: Shifted Erlang-2 vs.\ Poisson arrivals. Erlang-2 reduces RMSE by 47\% (12.24 vs.\ 23.04 vehicles). Ballistic integrator configuration.}
  \label{fig:arrival-comparison}
\end{figure}

On the other hand, the arrival process modeling has a substantial impact on lane-level
flow and fitness. We consider both a Poisson process and a shifted Erlang process for
vehicle arrivals, driven directly by trajectory-based PeMS data. The shifted Erlang process produces an overall average lane-level simulated flow that is approximately
\textbf{28$\%$ more accurate} when comparing the downstream sensor's 15-minute flow
measurements to their simulated counterparts. At Sensor~1, the shifted Erlang model
achieves an RMSE of 12.24 vehicles compared to 23.04 for Poisson, representing a
\textbf{47\% reduction in prediction error}, as illustrated in Figure.~\ref{fig:arrival-comparison}.
This represents a reduction in prediction
uncertainty by nearly one-half. We attribute this
improvement to both the use of staging and the $\tau$ parameter in the shifted Erlang model, which enforces a
minimum headway and provides the IDM with sufficient temporal spacing within each interval.
\enlargethispage{4\baselineskip}
  \vspace{-0.8em}

\subsection{Macro-Level Validation}
We evaluate macro-level simulation performance by assessing per-sensor agreement between simulated and empirical data. The simulated corridor spans 1.45~km and includes five mainline loop detector sensors across four lanes. Validation compares aggregated simulated and observed flow and speed at each sensor location using MAE, RMSE, NRMSE, and sMAPE as evaluation metrics.
\begin{table}[H]
    \centering
    \caption{Macro-Level Validation: Sensor Aggregates (Ballistic + Erlang2S).}
    \label{tab:macro-validation}
    \scriptsize
    \setlength{\tabcolsep}{6pt}
    \begin{minipage}{0.48\textwidth}
        \centering
        \textbf{(a) Flow (vehicles/15-min)}\\[0.5em]
        \begin{tabular}{crrrrr}
            \hline
            \textbf{Sensor} & \textbf{MAE} & \textbf{RMSE} & \textbf{Mean} & \textbf{NRMSE} & \textbf{sMAPE} \\
            \hline
            1 & 8.53 & 12.24 & 1168.79 & 1.05 & 3.36 \\
            2 & 15.23 & 19.55 & 1256.92 & 1.56 & 5.26 \\
            3 & 21.22 & 24.52 & 1264.75 & 1.94 & 7.40 \\
            4 & 24.68 & 28.47 & 1291.94 & 2.20 & 8.31 \\
            5 & 26.85 & 30.82 & 1280.10 & 2.41 & 9.08 \\
            \hline
            \textbf{Avg} & 19.30 & 23.12 & 1252.50 & \textbf{1.83} & 6.68 \\
            \hline
        \end{tabular}
    \end{minipage}
    \hfill
    \begin{minipage}{0.48\textwidth}
        \centering
        \textbf{(b) Speed (m/s)}\\[0.5em]
        \begin{tabular}{crrrrr}
            \hline
            \textbf{Sensor} & \textbf{MAE} & \textbf{RMSE} & \textbf{Mean} & \textbf{NRMSE} & \textbf{sMAPE} \\
            \hline
            1 & 3.28 & 3.85 & 23.79 & 16.16 & 18.38 \\
            2 & 3.61 & 4.20 & 24.03 & 17.48 & 20.33 \\
            3 & 3.43 & 3.92 & 24.37 & 16.08 & 18.73 \\
            4 & 3.31 & 3.92 & 24.64 & 15.90 & 17.73 \\
            5 & 2.56 & 3.51 & 25.33 & 13.87 & 13.53 \\
            \hline
            \textbf{Avg} & 3.24 & 3.88 & 24.43 & \textbf{15.90} & 17.74 \\
            \hline
        \end{tabular}
    \end{minipage}
    \begin{flushleft}
        \centering
        \footnotesize
        \textit{Note:} Fitness $= 0.5 \times 1.83 + 0.5 \times 15.90 = 8.87\%$ (NRMSE values shown as percentages).
    \end{flushleft}
\end{table}
\begin{figure}[H]
    \centering
    \includegraphics[width=0.92\textwidth]{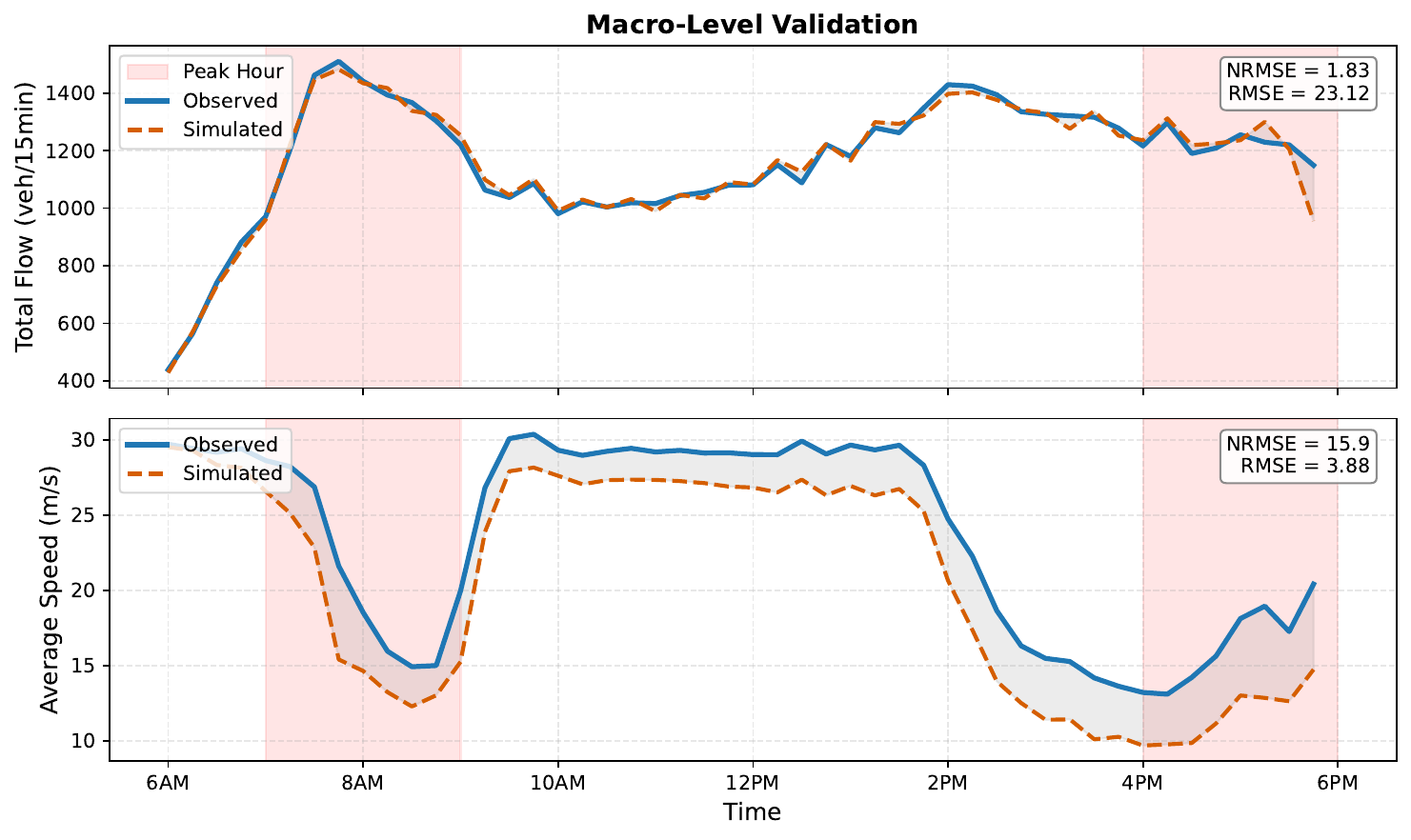}
    \vspace{-1.0em}
    \caption{Macro-level flow validation: observed PeMS vs. simulated (Ballistic + Erlang2S) at five mainline sensors. 15-minute intervals, 6:00 AM to 6:00 PM.}
    \label{fig:macro-validation}
\end{figure}
Our results show an average flow NRMSE of 1.83\% (MAE: 19.30, RMSE: 23.12) and speed NRMSE of 15.90\% (MAE: 3.24, RMSE: 3.88), yielding a combined fitness of 8.87 (Table~\ref{tab:macro-validation}).
The shifted Erlang-2 arrival process plays a significant role in accurately
reproducing total vehicle counts 62{,}445 vehicles recorded across the
five mainline loop detectors and two on-ramp loop detectors during the 12-hour study
period (6:00~AM to 6:00~PM). Figure~\ref{fig:macro-validation} demonstrates that our simulation model captures congestion onset and dissipation along the corridor. However, aggregated metrics can obscure lane-specific dynamics; we therefore present lane-level validation in the following subsection.

 \subsection{Micro-Level Validation: Lane Speed and Flow}
Aggregate validation provides a useful baseline, but it masks the lane-specific interactions that govern car-following behavior, merge conflicts, and congestion formation. To our knowledge, only \cite{11118854} has reported lane-level validation of simulated flow and speed against empirical detector data. Because driver decisions are made based on local lane conditions rather than corridor averages, validating simulation performance at the lane level is essential for assessing microscopic realism.
\begin{table}[htbp]
    \centering
    \caption{Lane-Level Micro Validation: Flow and Speed Accuracy.}
    \label{tab:micro-flow-speed}
    \scriptsize
    \setlength{\tabcolsep}{6pt}
    \begin{tabular}{cc|ccccc|ccccc}
        \hline
         & & \multicolumn{5}{c|}{\textbf{\large FLOW}}
             & \multicolumn{5}{c}{\textbf{\large SPEED}} \\
         & & \multicolumn{5}{c|}{\textbf{(vehicles / 15-min)}}
             & \multicolumn{5}{c}{\textbf{(m/s)}} \\
        \cline{3-12}
        \textbf{Sensor} & \textbf{Lane}
         & \textbf{MAE} & \textbf{RMSE} & \textbf{Mean} & \textbf{NRMSE} & \textbf{sMAPE}
         & \textbf{MAE} & \textbf{RMSE} & \textbf{Mean} & \textbf{NRMSE} & \textbf{sMAPE} \\
        \hline
        1 & 1 & 7.60 & 11.70 & 201.35 & 5.81 & 4.80 & 2.50 & 3.60 & 28.67 & 12.56 & 13.30 \\
          & 2 & 10.10 & 15.30 & 410.71 & 3.73 & 2.50 & 4.40 & 4.80 & 23.40 & 20.51 & 22.90 \\
          & 3 & 9.50 & 12.70 & 319.88 & 3.97 & 3.00 & 3.50 & 3.80 & 22.40 & 16.96 & 19.60 \\
          & 4 & 7.00 & 9.20 & 236.85 & 3.88 & 3.10 & 2.80 & 3.20 & 20.71 & 15.45 & 17.70 \\
        \hline
        2 & 1 & 10.80 & 15.10 & 210.10 & 7.19 & 6.00 & 2.70 & 3.90 & 29.19 & 13.36 & 13.70 \\
          & 2 & 14.30 & 19.40 & 417.92 & 4.64 & 3.60 & 4.40 & 4.70 & 25.01 & 18.79 & 22.60 \\
          & 3 & 24.30 & 28.70 & 345.58 & 8.30 & 7.30 & 3.80 & 4.20 & 18.67 & 22.50 & 21.30 \\
          & 4 & 11.50 & 15.00 & 283.31 & 5.29 & 4.10 & 3.60 & 4.00 & 23.05 & 17.35 & 23.80 \\
        \hline
        3 & 1 & 15.40 & 18.40 & 216.35 & 8.50 & 8.70 & 2.20 & 3.20 & 27.27 & 11.73 & 11.10 \\
          & 2 & 22.60 & 26.60 & 431.71 & 6.16 & 5.50 & 4.50 & 4.60 & 25.94 & 17.73 & 22.60 \\
          & 3 & 25.10 & 28.30 & 346.54 & 8.17 & 7.60 & 3.70 & 4.00 & 21.31 & 18.77 & 20.50 \\
          & 4 & 21.80 & 24.80 & 270.15 & 9.18 & 7.80 & 3.30 & 3.90 & 18.34 & 21.26 & 20.70 \\
        \hline
        4 & 1 & 18.60 & 22.70 & 220.17 & 10.31 & 9.80 & 2.40 & 3.60 & 28.90 & 12.46 & 12.20 \\
          & 2 & 26.90 & 31.60 & 437.67 & 7.22 & 6.50 & 5.00 & 5.20 & 24.71 & 21.05 & 24.90 \\
          & 3 & 28.40 & 31.50 & 349.21 & 9.02 & 8.60 & 3.70 & 4.40 & 23.69 & 18.57 & 20.80 \\
          & 4 & 24.90 & 28.10 & 284.90 & 9.86 & 8.40 & 2.20 & 2.60 & 21.26 & 12.23 & 13.10 \\
        \hline
        5 & 1 & 18.30 & 22.70 & 220.00 & 10.32 & 10.00 & 1.80 & 2.70 & 25.99 & 10.39 & 7.60 \\
          & 2 & 29.80 & 35.00 & 438.52 & 7.98 & 7.10 & 3.80 & 4.40 & 25.74 & 17.09 & 19.10 \\
          & 3 & 25.70 & 28.90 & 345.81 & 8.36 & 7.80 & 2.60 & 3.80 & 23.31 & 16.30 & 14.70 \\
          & 4 & 33.60 & 36.70 & 275.77 & 13.31 & 11.50 & 2.10 & 3.10 & 20.82 & 14.89 & 12.70 \\
        \hline
    \end{tabular}
      \vspace{0.5em}
    \begin{flushleft}
        \footnotesize
        \textit{Note:} Lane 1 = leftmost (fast lane); Lane 4 = rightmost (slow lane). NRMSE values shown as percentages.
    \end{flushleft}
\end{table}

\begin{figure}[H]
    \centering
    \includegraphics[width=0.90\textwidth]{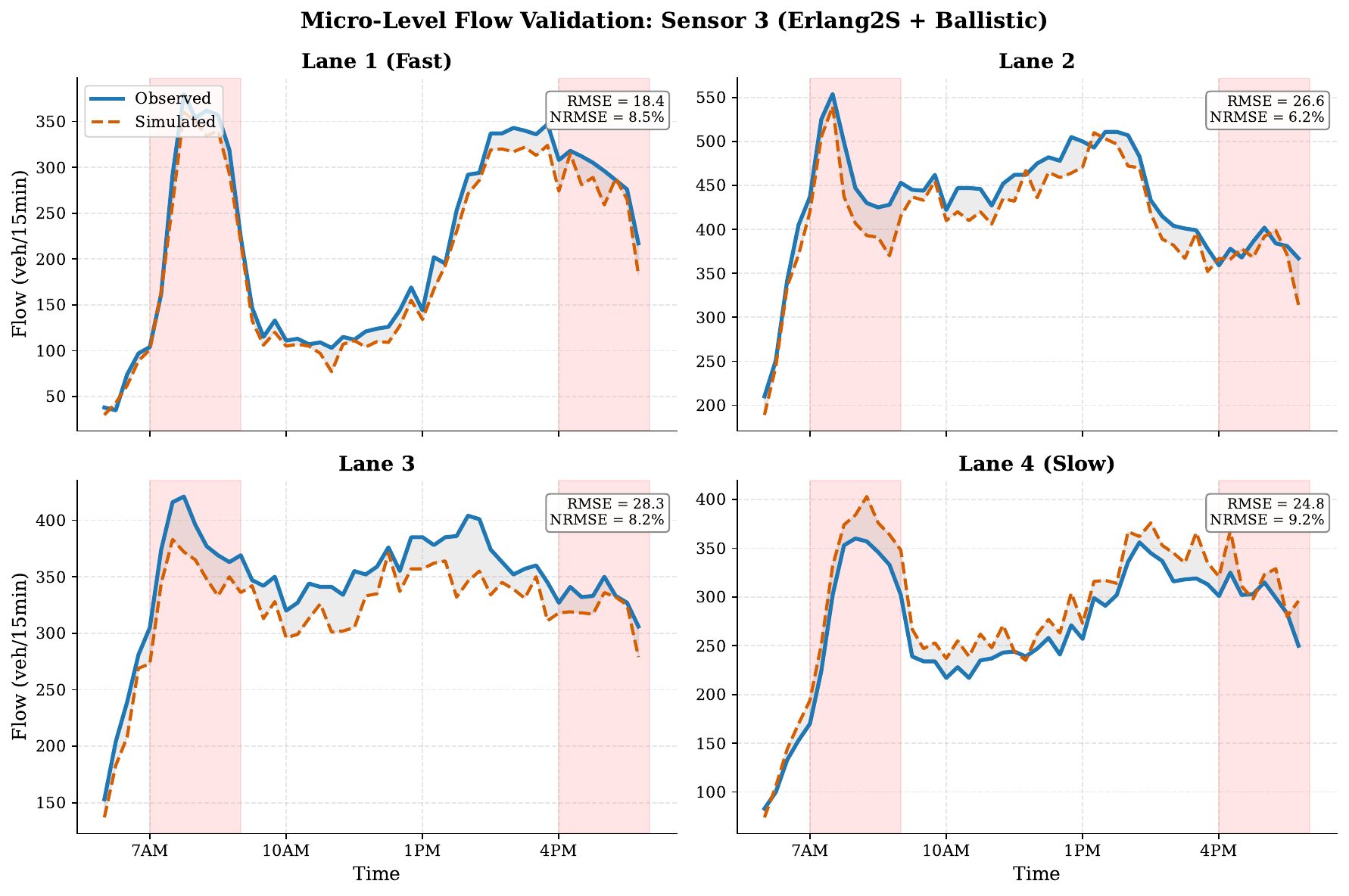}
    \caption{Lane-level flow validation at Sensor~3 (mid-corridor). L1 (fast lane, leftmost) shows highest accuracy (NRMSE = 8.5\%); L4 (slow lane, rightmost) exhibits greater variability due to on-ramp merging. Solid: observed; dashed: simulated. 15-minute intervals, 6:00 AM--6:00 PM.}
    \label{fig:micro-flow}
\end{figure}

\begin{figure}[H]
    \centering
    \includegraphics[width=0.90\textwidth]{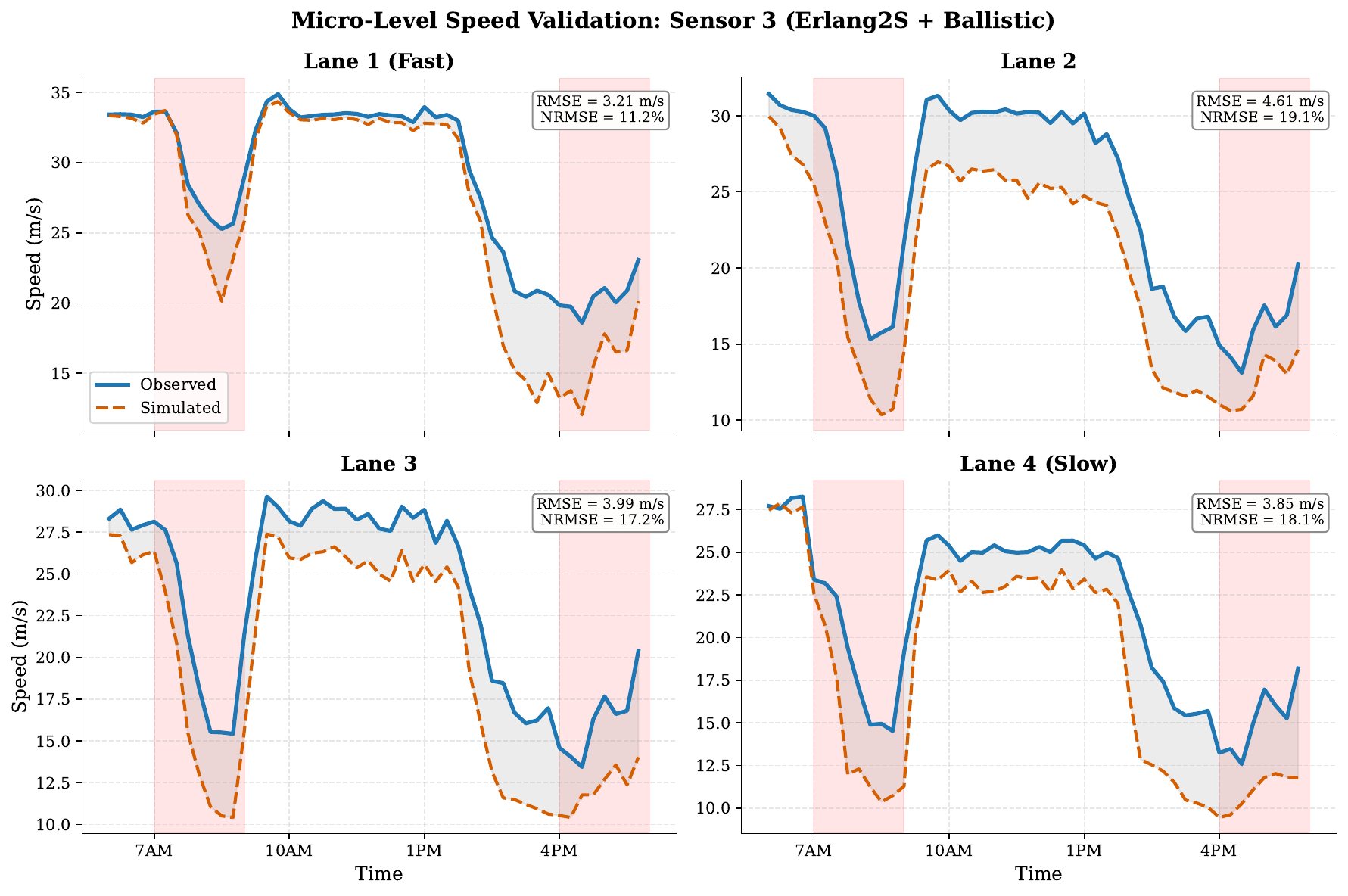}
    \caption{Lane-level speed validation at Sensor~3. The simulation captures general congestion trends but exhibits systematic bias in slower lanes (L3, L4) near merge zones. Speed NRMSE ranges from 11.7\% (L1) to 21.3\% (L4). Ballistic + Erlang2S configuration.}
    \label{fig:micro-speed}
\end{figure}

We validate our simulation by comparing observed and simulated 15-minute flow and speed measurements for each lane at each sensor. The results show clear and consistent lane-dependent patterns. Fast lanes exhibit the highest predictability, with flow NRMSE ranging from 5.8\% to 10.3\% across all five sensors, while slow lanes display increased variability associated with merging activity and heterogeneous driving behavior. Speed accuracy degrades at downstream sensors, where congestion and near stop-and-go dynamics dominate.
Table~\ref{tab:micro-flow-speed} reports complete lane-level diagnostics for all sensors, while Figures~\ref{fig:micro-flow} and~\ref{fig:micro-speed} illustrate representative per-lane behavior at Sensor~3, highlighting the model's ability to capture lane-specific congestion onset, persistence, and recovery.
\enlargethispage{4\baselineskip}
  \vspace{-0.8em}

\section{Discussion}
\label{sec:discussion}

The lane-level validation results presented in this study highlight a methodological gap in current microscopic traffic simulation practice. Although car-following models are inherently lane-based, validation is most often performed using aggregate measures such as total corridor flow and average speed. This mismatch between modeling resolution and validation granularity can conceal systematic errors that emerge at the lane level, particularly in congested or merging environments.
Reproducing lane-specific speed distributions, especially in slower lanes near merge regions, remains
challenging. Similar observations were reported by Gao et al.~\cite{11118854} for a single
bottleneck; our results confirm that such discrepancies persist across multiple sensors, suggesting
the issue is structural.

The negligible influence of numerical integration schemes on simulation accuracy is evident from our experiments: all eight integrators produced fitness values within 1\% of each other. The main difference is runtime: ballistic integration runs approximately 3$\times$ faster than higher-order methods like RK4. For microscopic traffic simulation, the simplest integrator is therefore sufficient. This integrator-insensitivity is consistent with IDM's smooth, differentiable ODE formulation and is not expected to generalize to piecewise models such as Gipps, which achieved comparable flow accuracy (NRMSE: 2.11\% vs.\ 2.12\%) but substantially degraded speed accuracy (17.23\% vs.\ 12.56\%), further validating the choice of IDM for lane-level speed validation.

In contrast, arrival process modeling strongly affects simulation accuracy. The shifted Erlang distribution enforces a minimum headway between vehicles, better matching how traffic actually enters a freeway and improving both aggregate and lane-level results. Lane-level analysis also reveals clear differences between lanes: fast lanes behave predictably, while slow lanes are heavily influenced by merging and stop-and-go traffic. These patterns are invisible when validation relies only on corridor-wide averages.

\section{Limitations and Future Directions}

We used standard IDM parameter values from literature since our primary
objective was evaluating the relative influence of integration schemes and
arrival processes on lane-level accuracy. Prior work has shown that
calibration outcomes depend on driving regime completeness
\cite{sharma2019more}. Future work will incorporate systematic
calibration of the IDM and related car-following models using established optimization
methods such as Simultaneous Perturbation Stochastic Approximation (SPSA), Genetic
Algorithms (GA), and Simulated Annealing. These methods will be tested within a controlled experimental framework to assess their impact on lane-level accuracy across multiple sensors. Future work will do multi-lane imputation to make the speed data for sensor 4 more realistic (currently it does single lane imputation). The current study is limited to a multi-lane freeway corridor. Future work will validate the framework on additional road types and a broader set of sensor stations to assess generalization.

\section{CONCLUSIONS}
\enlargethispage{3\baselineskip}
\label{sec:conclusions}
Our investigation yields three key findings. First, the choice of numerical integrator
has negligible impact on simulation accuracy for microscopic discrete-event traffic simulation. Across eight solvers, from ballistic kinematics to higher-order Runge-Kutta methods, variation in goodness-of-fit metrics remained below 1\%. The simplest and fastest method, ballistic integration, is therefore recommended.

Second, vehicle arrival process modeling has a substantially greater effect on simulation fidelity. The shifted Erlang-2 distribution improved overall fitness by approximately 28\% compared to Poisson arrivals. This result was consistent across all integrators, demonstrating that accurate arrival modeling deserves greater attention than numerical solver selection in calibration efforts.

Third, lane-level validation against empirical PeMS data reveals dynamics that aggregate metrics obscure. Using observations from five mainline loop-detectors across a four-lane freeway segment, we demonstrate that microscopic validation provides a more rigorous assessment of simulation fidelity than corridor-level metrics alone.

\appendix
\section{Appendices} \label{app:quadratic}
This snapshot is necessary because discrete-event simulation updates vehicles sequentially
\begin{algorithm}
\caption{Per-Vehicle State Update}
\begin{algorithmic}[1]
\Require vehicle $n$, time step $\Delta t$, solver $\mathcal{S}$
\Ensure updated position $x_n$, velocity $v_n$
\State Snapshot leader: $(\tilde{x}_p, \tilde{v}_p) \gets (x_p, v_p)$
\State Solve: $\mathbf{y}_1 \gets \mathcal{S}.\text{integrate}([v, a_{\text{IDM}}]^T, [x_n, v_n]^T, \Delta t)$
\State Extract: $x_n \gets \mathbf{y}_1[0]$, $v_n \gets \mathbf{y}_1[1]$
\State Clamp: $v_n \gets \max(0, \min(v_n, v_{\max}))$
\end{algorithmic}
\end{algorithm}

\bibliographystyle{scsproc}

\bibliography{assets/myReference}

\section*{Author Biographies}
\enlargethispage{7\baselineskip}
\textbf{\uppercase{Korede R. Bishi}} is a Ph.D. student in Computer Science at the University of Georgia, working in the Modeling, Simulation \& Analytics Lab (MSAL) under the supervision of Dr. John A. Miller. He received his B.S. in Computer Science from the University of Benin, Nigeria, in 2011. His research focuses on data-calibrated microscopic traffic simulation and agentic frameworks for autonomous simulation design and analysis using real-world sensor data. His email address is \email{korede.bishi@uga.edu}

\textbf{\uppercase{Casey Bowman}} is a Professor of Mathematics and Computer Science at University of North Georgia since 2007. He completed B.S. degrees in Mathematics and Computer Science in 2003, and an M.A. in Mathematics in 2005 from the University of Georgia. He received his Ph.D. in Computer Science from UGA in 2022. His research interests are microscopic traffic simulation, calibration of simulation models, agent-based simulation, and simulation optimization. His email address is \email{casey.bowman@ung.edu}.

\textbf{\uppercase{John A. Miller}} is a Professor in the School of Computing at the University of Georgia.  He has been the Associate Head for 6 years and has been Graduate Coordinator for 13 years.  His research interests include (i) Modeling \& Simulation, (ii) Web Services/Workflow, (iii) Database Systems, (iv) Big Data/Data Science, and (v) Bioinformatics.  Dr. Miller received a B.S. in Applied Mathematics from Northwestern University in 1980 and an M.S. and Ph.D. in Information and Computer Science from the Georgia Institute of Technology in 1982 and 1986, respectively.  As part of his co-operative undergraduate education program, he worked as a Software Developer at the Princeton Plasma Physics Laboratory.  In his areas of interest, Dr. Miller has authored over 200 research papers.  He is/has been an Associate Editor for the ACM Transactions on Modeling and Computer Simulation, SIMULATION: Transactions of the Society for Modeling and Simulation International, and IEEE Transactions on Systems, Man and Cybernetics, as well as an Editorial Board Member for the Journal of Simulation and International Journal of Simulation and Process Modelling.  He has also served as General/Program Chair for five research conferences: ANSS, WSC, ICWS and SCC, and ACM-SE.  He has worked on funded research in the areas of workflow, Web services and the Semantic Web with Dr. Budak Arpinar for several years and nine years of funded research with Dr. Jessica Kissinger on the development of databases for eukaryotic pathogens with access support via Web services and data integration based on Semantic Web technologies including OWL ontologies.  His most recent work has focused on the development of knowledge-enhanced deep learning for pandemic preparedness and prediction. His email address is \email{jamill@uga.edu}

\end{document}